# Analyzing whether workplace smoking bans can reduce the probability of smoking?

Tianjiao He

June 21, 2021

# Abstract

The rapid increase of smoking-related diseases and deaths globally is driving us to find an effective approach to reduce the smoking rate. This study aims to determine whether indoor smoking bans at workplaces can effectively reduce the smoking rate. The "Smokeban" dataset used for this study is an observational dataset that contains some socio-demographic factors, whether people smoke, and whether smoking bans exist. Since the observational data used in the study did not randomize people into "with smoking bans" group and "without smoking bans" group, confounders may cause bias in the estimation of whether the smoking bans can reduce smoking rates. The propensity score matching (PSM) method can reduce these biases via using a logistic regression model to predict the similarities of people in those 2 groups and using the nearest neighbor matching technique to match people who are the most similar. After reducing the bias, another regression model was created to interpret the relationship between the probability of smoking and the indoor smoking bans. We conclude by arguing that with the existence of indoor smoking bans, the probability of people who smoke can be decreased greatly.

# Introduction

Tobacco use is the leading cause of preventable disease and death worldwide. (Teresa Janz, 2012) Each year, 7 million people die because of smoking-related diseases like cancer, heart disease, stroke, and lung diseases. If the pattern of smoking all over the globe does not change, more than 8 million people a year will die from diseases related to tobacco use by 2030. (Centers for Disease Control and Prevention, 2021) Not only does smoking affect smokers' health, but it also influences people around them. Breathing second-hand smoke can have immediate adverse effects on your blood and blood vessels, increasing the risk of having a heart attack. (Centers for Disease Control and Prevention, 2020) Given these facts, it is important and urgent to find a way to reduce the smoking rate and prevent people from second-hand smoke. In Canada, the federal government has banned tobacco use in workplaces (including restaurants, bars, and casinos), by all territories and provinces by 2010. (Wikipedia, 2018) Therefore, we want to figure out whether smoking bans in workplaces can control tobacco use.

This study will focus on determining whether people can reduce smoking if smoking bans exist in workplaces. The study will take advantage of an observational dataset("SmokeBan") that includes whether people smoke in the presence of smoking bans, as well as some socio-demographic factors such as age, gender, education level, and race. Since the "SmokeBan" dataset did not randomly assign people to groups with or without indoor smoking bans, there might be confounders that influence the relationship between smoking rate and smoking bans. Propensity score matching is a statistical matching method that attempts to estimate the effect of a treatment (whether smoking bans existed) on the outcome of interest (smoking rate) by accounting for the covariates (socio-demographic factors) that predict receiving the treatment. (Wikipedia, 2021) PSM can reduce the bias caused by confounders when estimating the treatment effect obtained from comparing outcomes among units that received the treatment versus those that did not. (Wikipedia, 2021) Further analysis of PSM will be included in this study to find the relationship between smoking rate and smoking bans.

Based on the research from the U.S. Department of Health and Human Services, smoking bans can prevent initiation of tobacco uses among youth, increase cessation, and reduce smoking prevalence among workers and the general population. (U.S. Department of Health and Human Services, 2020) Hence, we hypothesize that the existence of a smoking ban can reduce the possibility of smoking.

The following sections will go further in-depth on the dataset we selected, which include the data cleaning process, the PSM method used to analyze the relationship between smoking rate and smoking bans, the results of the method, as well as the conclusions to our hypothesis.

# Data

## Data Collection Process

The dataset "SmokeBan" is free, open observational data from this websitehttps://vincentarelbundock.github.io/Rdatasets/datasets.html. The data were downloaded into a .csv format and was loaded into R using the read_csv() function.

The "Smokeban" dataset is cross-sectional with 10,000 observations, created by Professor William Evans of the University of Maryland. The dataset was cleaned from an 18,090 observational dataset collected in the National Health Interview Survey in 1991 and then again (with different respondents) in 1993. (RDocumentation, 2020)

There are some limitations of this dataset. It was collected over 30 years ago and the data only represents the situation in the past. However, since it is hard to find open, free observational data with a sample size that is big enough to analyze the relationship between smoking rate and smoking bans, I decided to use this dataset for further analysis.

## Data Summary

This dataset contains information on whether the individuals smoked, whether a smoking ban existed at the workplace, and other individual characteristics. Table 4 in the Appendix(Section 1) shows the first 6 rows of the SmokeBan dataset. There are 7 variables in this dataset, which are introduced in the following bullet points.

The response variable in this dataset consisted of:
- whether the individual is a current smoker?

The categorical predictor variables consisted of:
- whether a work area smoking ban exists or not
- highest education level attained
- whether the individual is African-American?
- whether the individual is Hispanic?
- gender

The numerical predictor variable includes:
- age

We observed that asking whether a person is African-American or Hispanic may lead to racial discrimination. I wanted to replace these two questions with "whether an individual belongs to minorities". However, people who answer "yes" to either of the two questions are easily classified as ethnic minorities, but those who answer "no" to both questions may or may not belong to minorities. So, it is not reasonable to replace those two questions, and I have to keep them.

To clean this dataset, I used "drop_na" function to remove null values in the dataset. But there is no missing value in the 10,000 observations, so the quality of the dataset is quite good. For the categorical variables of smoked or not, the values of "yes" and "no" are turned into numerical values where 1 indicates "yes" and 0 indicates "no" using rename function. For whether the indoor smoking bans existed, since "no smoking bans" is defined as treated, the value of "no" is turned into 1, which indicates the individual belongs to the treatment group. The value of "yes" is turned into 0, which indicates the individual belongs to the control group. To be more specific, the treatment group represents people whose indoor workplaces lacked smoking bans, and the control group represents people whose workplaces had smoking bans. The binary numerical values allow us to use the logistic regression model to find the probability of being treated based on whether the smoking bans existed.

All the 7 variables mentioned before will be needed in the propensity score matching method. Therefore, I will introduce all of them in detail here.

Table 1: Summary of Important Variables

| Variable | Categories | Number of observations in each category |
| --- | --- | --- |
| Smoking Ban | Without smoking bans, With smoking bans | 6098, 3902 |
| Smoker | Smoker, Non-smoker | 2423, 7577 |
| Gender | Female, Male | 5637, 4363 |
| African American | African American, Non African American | 769, 9231 |
| Hispanic | Hispanic, Non Hispanic | 1134, 8866 |
| Education level | High school, High school(drop out), College(drop out), College, Master | 3266, 912, 2802, 1972, 1048 |

Table 1 shows the summary of categorical variables. Among the 10000 observations, 3902 individuals worked without indoor smoking bans while 6098 workers with them. The sample sizes for "with smoking bans" and "without smoking bans" are reasonably large. We observed that there are only two categories in gender. It is not appropriate when considering gender diversity. Maybe when the data was collected 30 years ago, gender diversity was not taken seriously. It is better to use sex rather than gender here.

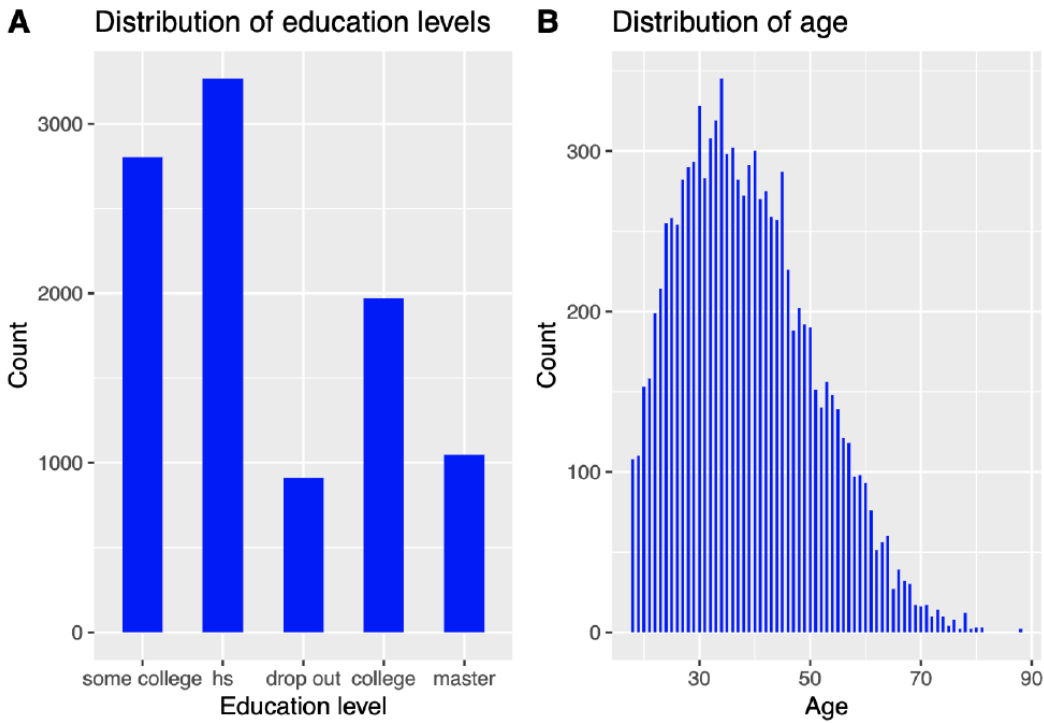

Barplot A above shows the distribution of the highest education level attained for each observation. "hs", "drop out", "some college", "college", "master" represent "high school graduate", "high school drop out", "college drop out", "college graduate", "master's degree(or higher)". From the plot, we find that nearly 60% of people were graduated from high school or had some college experience, while only 30% of them graduated from college or got a master's degree or higher. We need this variable when predicting the probability of being assigned to the treatment group where there is no smoking ban at the workplace. We think the education level can influence people's job, and thus influencing whether smoking bans exists at workplace in some way.

Barplot B above shows the distribution of age for this dataset. The age ranges from 18 to 88. The average age is 38.69 and the standard deviation is 12.11. The distribution of age is right-skewed where most people are within 25 to 50. This variable is also used to predict the propensity score so that we can divide the data into treatment and control groups using the nearest neighbor matching approach.
All analysis for this report was programmed using R version 4.0.4.

# Methods

The propensity score matching method was introduced by Paul R. Rosenbaum and Donald Rubin in 1983. (Wikipedia, 2021) It relies on the assumption that conditional on some observable characteristics, untreated units in the control group can be compared to treated units in the treatment group, as if the treatment has been fully randomized. (The World Bank, n.d.) In this way, PSM seeks to mimic randomization to overcome issues of selection bias that plague non-experimental methods such as observational studies. (The World Bank, n.d.) PSM also requires a large sample size on both treatment and control units. Further, if unobservable characteristics exist between the treated and untreated units, PSM will provide biased estimates. Therefore, all the relevant characteristics related to treatment assignment and outcome of interest should be selected.

Propensity score matching consists of several analytic steps:
1. Based on the covariates(age, gender, education level, African American, Hispanic), we can estimate the propensity score for each observation. A logistic regression model is used where the treatment assignment(lacking indoor smoking bans) is the outcome variable and the covariates we think that can explain the treatment assignment are predictor variables. The model allows us to compare the similarities between observations. We then use this model to predict a propensity score for each observation in the dataset.

   Formula (1) shows the logistic regression model to calculate the propensity score.

   $$propensity = log(\frac{p(t)}{1-p(t)}) = \beta_0 + \beta_1 x_{education(hs)} + \beta_2 x_{education(hsdropout)} + \beta_3 x_{education(master)}$$

   $$+ \beta_4 x_{education(collegedropout)} + \beta_5 x_{africanamerican(yes)} + \beta_6 x_{hispanic(yes)} + \beta_7 x_{gender(male)} + \beta_8 x_{age} \quad (1)$$

   In formula (1), $p(t)$ is the probability of an observation being treated(no smoking bans). $\frac{p(t)}{1-p(t)}$ is called odds. We take the natural log of odds as the propensity score. $\beta_0$ is the intercept term. $\beta_1, \beta_2, \beta_2, ... \beta_7$ are the coefficients of categorical variables(education, African American, Hispanic and gender). It should be interpreted as the change in log odds for being in different categories. For example, $\beta_1$ represents the change in log odds when the education level is high school rather than college instead while fixing the other variables. We noticed that the education level of college does not exist in formula (1). Because we define all the x variables related to education to be 0 to represent the education level of college. The interpretation for other categorical variables is the same as "education".

2. After getting the propensity score, the nearest neighbor matching approach is used to match each treated participant to a single untreated participant with the most similar propensity score. We use matchit() function to implement this 1-to-1 nearest neighbour matching. (Gary King et al. 2019) The arguments of matchit() function include method="nearest", ratio=1, and the formula where treatment is the outcome, and covariates are predictor variables. If the number of treated participants is less than the untreated participants, the extra untreated participants that are not matched with any treated participant will be disregarded.

3. After the matching procedure, we need to evaluate the quality of matching. We want to use both quantitative and qualitative evaluation.

   First, we use a graphical approach(qualitative evaluation) in the package of Hmisc which was created by Harrel in 2015. (Harrell, 2015) It can draw a back-to-back histogram that evaluates the distributional similarity between the treatment and control groups using histbackback(). The arguments of the function are the propensity score and the treatment assignment for each observation. If the shapes of the 2 sets of histograms are similar, it indicates the propensity scores are distributed similarly among the 2 groups, and thus reducing imbalances of the covariates. (Antonio Olmos, 2015) We compare the histograms before and after matching. If there is a great improvement after matching, then the quality of matching is good.

Secondly, we use a quantitative evaluation. The plot.summary.matchit() function in R packages can generate a dot plot with variable names on the y-axis and standardized mean differences on the x-axis. (Antonio Olmos, 2015) The argument of the function is the result of matchit() function. Each point in the plot represents the standardized mean difference of the corresponding covariate in the matched or unmatched observations. (Rdocumentation, 2021) It is a simple way to display covariate balance before and after matching. If the standardized mean differences of covariates are close to 0 after matching, then our matching technique is appropriate. (Antonio Olmos, 2015) The formula for the standardized mean difference(SMD) is calculated in formula (5) in the Appendix(Section 2). (Zhongheng Zhang et al. 2019)

If the evaluations show that the matching is not good enough, we may need to consider whether the covariates selected in step 1 are appropriate for explaining the treatment assignment or not.

4. In the final step, we need to evaluate whether the presence of smoking bans reduces smoking while the covariates are balanced. Since our outcome of interest is binary(whether people still smoke), we used a logistic regression model where the response variable is "whether people smoke" and the predictor variables include covariates and the treatment variable. The model should be built based on the matched observations which we got in step 3 to ensure that biases for confounders were reduced.

Formula (2) shows the logistic regression model to evaluate the effect of indoor smoking bans on the smoking rate.

$$log(\frac{p(smoke)}{1-p(smoke)}) = \beta_0 + \beta_1 x_{treat} + \beta_2 x_{age} + \beta_3 x_{education(hs)} + \beta_4 x_{education(hsdropout)} + \beta_5 x_{education(master)}$$
$$+ \beta_6 x_{education(collegedropout)} + \beta_7 x_{africanamerican(yes)} + \beta_8 x_{hispanic(yes)} + \beta_9 x_{gender(male)} \quad (2)$$

In formula (2), $p(smoke)$ is the probability of an individual still being a smoker. $\beta_0$ is the intercept term. $\beta_1$ is the coefficient of the variable for treatment. When $x_{treat}$ equals 1, it indicates the individual is in the treatment group where work area smoking bans did not exist. On the contrary, if $x_{treat}$ equals 0, the individual is in the control group with work area smoking bans existing. So, $\beta_1$ the change in log odds between being treated and not being treated while other variables are fixed.

$\beta_5$ is the slope for numerical variable age. It represents the change in log odds with one unit of increase in age while fixing other variables. $\beta_3, \beta_4, \beta_5, ...\beta_9$ are the coefficients of categorical variables, which the change in log odds for being in different categories. Apart from using this model to evaluate the treatment effect on the outcome of interest, we can use this model to predict the probability of whether people smoke based on the characteristics included in this model.

# Results

Our original goal is to analyze whether indoor smoking bans could reduce smoking. I will show the result of the propensity score matching method in steps.

**Step 1: Estimate the propensity score using a logistic regression model**

I selected 5 covariates (age, gender, education level, African American, Hispanic) that may explain whether people are treated or not. The 5 covariates may influence people's jobs, and thus affecting the existence of indoor smoking bans. I used a logistic regression model to find the relationship between the existence of indoor smoking bans and the 5 covariates.

Table 2. Summary of logistic regression model

| Characteristic | log(OR)[1] | 95% CI[1] | p-value |
|---|---|---|---|
| age | -0.01 | -0.01, 0.00 | <0.001 |
| education | | | |
| college | — | — | |
| hs | 0.67 | 0.55, 0.79 | <0.001 |
| hs drop out | 1.0 | 0.83, 1.2 | <0.001 |
| master | -0.14 | -0.30, 0.03 | 0.11 |
| some college | 0.36 | 0.24, 0.48 | <0.001 |
| afam | | | |
| no | — | — | |
| yes | -0.10 | -0.26, 0.05 | 0.2 |
| hispanic | | | |
| no | — | — | |
| yes | -0.08 | -0.22, 0.06 | 0.3 |
| gender | | | |
| female | — | — | |
| male | 0.51 | 0.43, 0.60 | <0.001 |

[1] OR = Odds Ratio, CI = Confidence Interval

Table 2 shows the summary of the logistic regression model. We can observe that the p-values for the coefficients of "age", "hs"(high school), "hs drop out"(high school drop out), "some college" and "male" are statistically significant(<0.05). It indicates that age, highest education level attained and gender all have a strong relationship with the existence of smoking bans. Some authors (Austin, Grootendorst & Anderson, 2007; Caliendo & Kopeinig, 2008) suggested that the model should include not only statistically significant variables but also variables known to be associated. (Antonio Olmos, 2015) Therefore, we keep the variable related to race("afam"(African American), "hispanic"(Hispanic)) since their p-values are not extremely large($\leqslant 0.3$).

Based on the summary of the model, we show the logistic regression model to calculate the propensity score in formula (3).

$$propensity = log(\frac{p(t)}{1-p(t)}) = -0.81 + 0.67 x_{education(hs)} + 1.0 x_{education(hsdropout)} - 0.14 x_{education(master)}$$

$$+0.36 x_{education(collegedropout)} - 0.10 x_{africanamerican(yes)} - 0.08 x_{hispanic(yes)} + 0.51 x_{gender(male)} - 0.01 x_{age} \quad (3)$$

The interpretation of the coefficients in formula (3) is in the Appendix(Section 3). With the formula (3), we can calculate the probability of each observation being treated by using formula (6) in the Appendix(Section 4) to convert log odds to probabilities. To avoid more computation, we can just use the log odds as propensity score since while the log odds increase or decrease, the probability of being treated also increases or decreases accordingly.

**Step 2: Implement the nearest neighbor matching technique**

In our study, the treatment group is people whose workplaces lacked smoking bans. There are 3902 observations in the treatment group and 6098 observations in the control group. Since there are 3902 observations in the treatment group, we need to find 3902 matches in the control group with the most similar propensity score. 2196 observations that were not matched in the control group will be discarded since we used 1-to-1 matching. The sample sizes for both the treatment and the control group are reasonably large, which satisfies the requirement of propensity score matching.

We can visualize the distribution of propensity scores of those who were matched using plot() with type

= "jitter". The argument of plot() is the matching result of matchit() we described in the methods section.

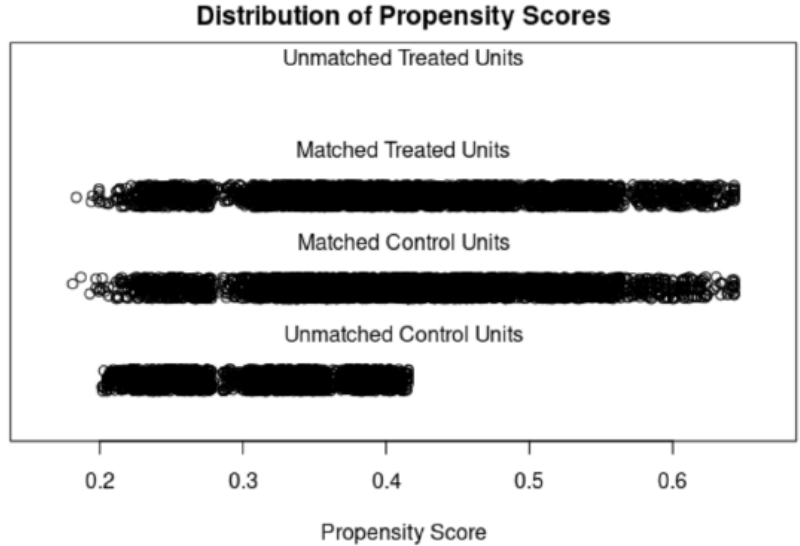

Figure 1: Distribution of Propensity Scores

In Figure 1, we observe that the 2196 unmatched observations have propensity scores within 0.2 to 0.4. It indicates that there are other observations in the control group whose propensity scores are more similar to the units in the treatment groups. The distributions of propensity scores for the matched units in the treatment group and the control group are very similar, which interprets the matching approach always chooses the most similar propensity score.

**Step 3: Evaluate the matching technique**

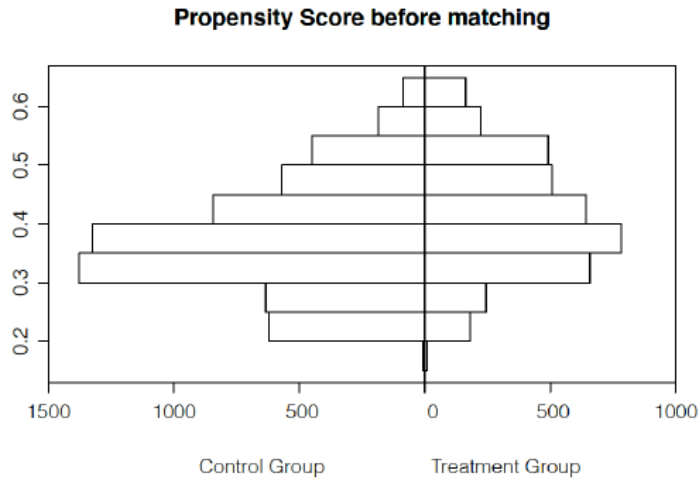

Figure 2: Propensity Score before matching

Figure 2 shows the distribution of propensity score before matching. The left histogram shows the distribution of propensity scores for the control group, and the right histogram is for the treatment group. We can observe that the shape of 2 sets of histograms differs a lot, indicating the imbalances for covariates before the matching technique.

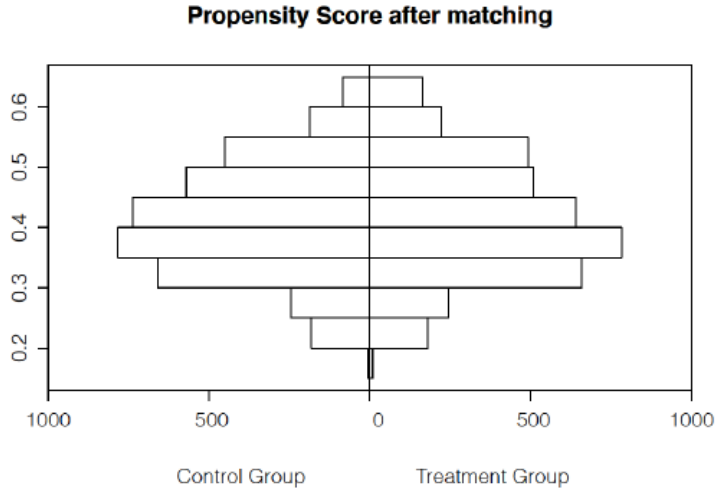

Figure 3: Propensity Score after matching

Figure 3 shows the distribution of propensity score after matching. We observe that there is a remarkable improvement in the match between the two distributions of propensity scores. It indicates the two groups are much more similar in terms of their propensity scores. So, the quality of matching is good. Then we evaluate the matching technique by the standardized mean differences of covariates.

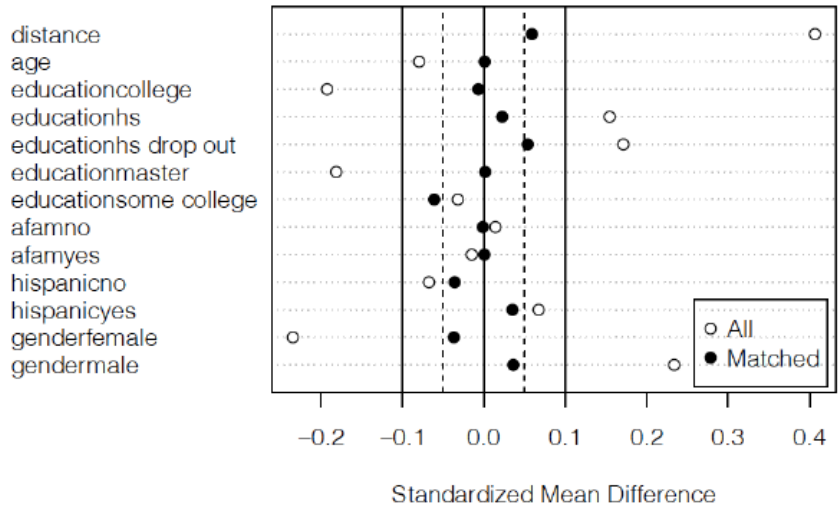

Figure 4: The standardized mean differences of covariates

In figure 4, the black dots represent the standardized mean difference of the corresponding covariates after matching. The white dots represent the standardized mean difference before matching. The y-axis is variable names, and categorical variables are followed by the corresponding categories. For example, "afamno" and "afamyes" represent "is African American" and "non African American". It is clear to see that the balance was quite poor before matching, but nearest neighbor matching improved balance on all covariates. The standardized mean differences of all covariates are close to 0, and thus the matching technique is considered good enough. (Antonio Olmos, 2015)

In both the quantitative and qualitative evaluations, we observed that the 5 covariates were balanced after matching. Therefore, the quality of matching is good, and the bias of confounders can be reduced when predicting the effect of treatment on the outcome of interest.

**Step 4: Evaluate outcomes using another logistic regression model with the matched data**

After matching, the balanced data were used in a logistic regression model to evaluate whether the existence of smoking bans could reduce smoking. Table 2 shows the summary of the result of the logistic regression model.

Table 3: Logistic Model Results for Outcome of Interest

| Table | Estimate | SE | z-value | p-value | statistical significance |
|---|---|---|---|---|---|
| Intercept | -1.647 | 0.122 | -13.502 | < 2e-16 | *** |
| Treatment | 0.262 | 0.053 | 4.961 | 7.02e-07 | *** |
| Age | -0.009 | 0.002 | -4.097 | 4.19e-05 | *** |
| Education_high_school | 1.081 | 0.0911 | 11.868 | < 2e-16 | *** |
| Education_high_school_drop_out | 1.485 | 0.112 | 13.315 | < 2e-16 | *** |
| Education_master | -0.498 | 0.163 | -3.06 | 0.002 | ** |
| Education_some_college | 0.69 | 0.095 | 7.252 | 4.11e-13 | *** |
| African_American_yes | -0.129 | 0.1 | -1.287 | 0.198 | |
| Hispanic_yes | -0.595 | 0.089 | -6.662 | 2.70e-11 | *** |
| Gender_male | 0.2 | 0.053 | 3.765 | 1.66e-4 | *** |

From table 3, we observe that the p-value for the treatment variable is statistically significant(<0.05). Therefore, the treatment indeed has effects on the outcome of interest. To further analyze whether the existence of smoking bans reduces the probability for people to smoke, we write down the formula for the logistic regression model.

$$log(\frac{p(s)}{1-p(s)}) = -1.65 + 0.26x_{treat} + 0.01x_{age} + 1.08x_{education(hs)} + 1.49x_{education(hsdropout)}$$
$$- 0.50x_{education(master)} + 0.69x_{education(collegedropout)} - 0.13x_{africanamerican(yes)}$$
$$- 0.60x_{hispanic(yes)} + 0.2x_{gender(male)} - 0.60x_{hispanic(yes)} + 0.2x_{gender(male)} \quad (4)$$

Formula (4) shows the logistic regression model to evaluate the effect of indoor smoking bans on the probability of smoking. p(s) is the probability of an individual who still smokes. The coefficient for the variable of treatment is 0.26. When $x_{treat}$ equals 1, it indicates the individual is in the treatment group where work area smoking bans did not exist. When $x_{treat}$ equals 0, the individual is in the control group with work area smoking bans existing. So, when the indoor working area lacked smoking bans(treated), there is a 0.26 increase in log odds compared with smoking bans existing (not treated) while keeping other explanatory variables fixed. It is equivalent to say that there is a 0.26 decrease in log odds if the smoking bans existed compared with no smoking bans.

Using the log converting formula (6) in the Appendix(Section 4), we can have the probability p = 0.5652. Therefore, we can conclude that the probability of smoking for individuals would decrease 56.5% if there exist smoking bans at workplaces. It satisfies our hypothesis at the very beginning.

# Conclusions

In this study, our hypothesis is the probability of smoking can be reduced with the existence of smoking bans. The Propensity Score Matching method was applied to adjust the treatment effect for measured confounders in the observational dataset("SmokeBan"). This assured that the only difference between the treatment group and control group is whether the smoking bans existed.

We used a logistic regression model to predict the similarity of observations in treatment and control groups based on people's age, education level, gender, and race. Then, with the help of the nearest neighbor matching technique, we could match each individual in the treatment group with the most similar one in the control group. The matching technique reduced the biases for measured confounders based on the qualitative and quantitative methods we introduced in the Methods section. Finally, another logistic regression model was used to estimate the effect of smoking bans on the smoking rate. The result shows that the probability of smoking for individuals will decrease 56.5% if smoking bans exist at workplaces. Therefore, our hypothesis is satisfied according to this study.

The result of this study shows that the ban on indoor smoking is very effective in reducing smoking. Therefore, Canada's legislation against indoor smoking is conducive to reduce smoking-related diseases and deaths. We should vigorously promote the ban on indoor smoking globally to reduce the number of people who die of smoking every year.

## Weaknesses

In this study, there are weaknesses to improve in the future. First, the dataset I used is outdated. There are also some ethical issues regarding race and gender that were discussed in the Data section. Besides, the dataset only contains 7 variables, so I cannot select other covariates in predicting the propensity score. Characteristics such as working environment, whether employed or not and income may increase the credibility of the propensity score analysis since the selection of proper covariates directly determines the quality of propensity score estimation. Apart from that, there are some flaws with the propensity score matching method itself. It cannot match on unobserved confounders. For example, working environments may affect both the existence of indoor smoking bans and the smoking rate. Hence, I had included this variable in our study, the quality of propensity score analysis would have been better. In addition, when selecting the covariates for explaining the treatment assignment, I did not use model selection methods to determine which set of covariates is more appropriate. I can improve this in the future.

## Next Steps

In the next steps, I want to implement the following tasks in the future. For the data, I will try to find some latest datasets that contain more variables with no ethical issues. If there was no appropriate dataset, I could implement a survey to satisfy all the requirements. For the unobserved confounders, Pearl found that confounders can be encoded into a directed acyclic graph, which can then, in turn, be used by the researcher to select specific confounders for the estimation of the propensity score. (Pearl, 2000) I may learn this method as my next step. Finally, for the model selection method, I find that the Akaike information criterion (AIC) is a way to estimate the quality of each model with different sets of predictor variables. (McElreath, 2016) It is also an interesting method to learn in the future.

## Discussion

In this study, we discussed the effect of indoor smoking bans on the probability of whether people smoke. We took advantage of an observational dataset("SmokeBan) that contains some socio-demographic factors, whether people smoke, and whether smoking bans exist. The propensity score matching method was used to make causal inferences in our observational studies. Using this method, we reduced the biases of the confounding variables, so we can analyze the relationship between indoor smoking bans and the smoking rate without confounding. Finally, we reached the conclusion that the existence of workplace smoking bans can reduce the probability of smoking by 56.5%. Hence, we should call for setting indoor bans on smoking globally.

# Bibliography

1. Teresa Janz. (2012) Current smoking trends. Statistics Canada Catalogue No. 82-624-X. https://www150.statcan.gc.ca/n1/pub/82-624-x/2012001/article/11676-eng.htm
2. Centers for Disease Control and Prevention. (2021) Smoking & Tobacco Use. Office on Smoking and Health, National Center for Chronic Disease Prevention and Health Promotion. https://www.cdc.gov/tobacco/data_statistics/fact_sheets/fast_facts/index.htm
3. Centers for Disease Control and Prevention. (2020, February 27) Health Effects of Secondhand Smoke. Office on Smoking and Health, National Center for Chronic Disease Prevention and Health Promotion. https://www.cdc.gov/tobacco/data_statistics/fact_sheets/secondhand_smoke/health_effects/index.htm
4. Wikipedia (2018, October) Smoking in Canada. Wikipedia. https://en.wikipedia.org/wiki/Smoking_in_Canada (Last Accessed: 18 June 2021)
5. Wikipedia (2021, May 22) Propensity score matching. Wikipedia. https://en.wikipedia.org/wiki/Propensity_score_matching
6. U.S. Department of Health and Human Services. (2014) The Health Consequences of Smoking—50 Years of Progress: A Report of the Surgeon General. U.S. Department of Health and Human Services, Centers for Disease Control and Prevention, National Center for Chronic Disease Prevention and Health Promotion, Office on Smoking and Health. (Last Accessed: 8 September 2020)
7. RDocumentation. (2020, February 6) SmokeBan: Do Workplace Smoking Bans Reduce Smoking? RDocumentation, AER (version 1.2-9) https://www.rdocumentation.org/packages/AER/versions/1.2-9/topics/SmokeBan
8. The World Bank. (n.d.) Propensity Score Matching. The World Bank Group. https://dimewiki.worldbank.org/Propensity_Score_Matching
9. Gary King and Richard Nielsen. (2019) Why Propensity Scores Should Not Be Used for Matching. Political Analysis, 27, 4, Pp. 435-454. Publisher's Version Copy at https://j.mp/2ovYGsW Harrell, F. E. (2015). Hmisc: Harrell Miscellaneous. R package version 3.15-0
10. Antonio Olmos, Priyalatha Govindasamy (2015) Propensity Scores: A Practical Introduction Using R. Journal of MultiDisciplinary Evaluation Volume 11, Issue 25
11. RDocumentation. (2021, May 26) plot.summary.matchit: Generate a Love Plot of Standardized Mean Differences. RDocumentation, MatchIt (version 4.2.0) https://www.rdocumentation.org/packages/MatchIt/versions/4.2.0/topics/plot.summary.matchit
12. Austin, P. C., Grootendorst, P., & Anderson, G. M. (2007) A comparison of the ability of different propensity score models to balance measured variables between treated and untreated subjects: A Monte Carlo study. Statistics in Medicine https://rdrr.io/cran/MatchIt/man/plot.summary.matchit.html
13. Caliendo, M., & Kopeinig, S. (2008). Some practical guidance for the implementation of propensity score matching. Journal of Economic Surveys
14. Pearl, J. (2000). Causality: models, reasoning, and inference. Cambridge University Press: New York.
15. McElreath, Richard (2016). Statistical Rethinking: A Bayesian Course with Examples in R and Stan. CRC Press.
16. Zhongheng Zhang, Hwa Jung Kim, Guillaume Lonjon, Yibing Zhu. (2019). Balance diagnostics after propensity score matching. AME Big-Data Clinical Trial Collaborative Group https://www.ncbi.nlm.nih.gov/pmc/articles/PMC6351359/

# Appendix

**Section 1: A glimpse of the data from "SmokeBan" dataset**

Table 4. SmokeBan Dataset

| smoker | ban | age | education | afam | hispanic | gender |
|--------|-----|-----|-----------|------|----------|--------|
| yes | yes | 41 | hs | no | no | female |
| yes | yes | 44 | some college | no | no | female |
| no | no | 19 | some college | no | no | female |
| yes | no | 29 | hs | no | no | female |
| no | yes | 28 | some college | no | no | female |
| no | no | 40 | some college | no | no | male |

Table 4 shows the first 6 rows of the "SmokeBan" dataset. There are 7 variables in this dataset. The meaning for each variable is shown in the following bullet points.
- smoker: whether the individual is a current smoker?
- ban: whether a work area smoking ban exists or not
- age: age of the individual
- education: highest education level attained
- afam: whether the individual is African-American?
- hispanic: whether the individual is Hispanic?
- gender: gender of the individual

**Section 2: The formula for the standardized mean difference(SMD)**

$$SMD = \frac{\bar{X}_1 - \bar{X}_2}{\sqrt{(s1^2 + s2^2)/2}} \quad (5)$$

In formula (5), $\bar{X}_1$ and $\bar{X}_2$ are the sample means for the treatment group and the control group. $s_1^2$, $s_2^2$ are the sample variance for the treatment group and the control group. If the standardized mean differences of covariates are close to 0 after matching, it indicates that people in the two groups are similar considering all the covariates.

**Section 3: The interpretation of the coefficients in formula (3)**

$$propensity = \log(\frac{p(t)}{1-p(t)}) = -0.81 + 0.67 x_{education(hs)} + 1.0 x_{education(hsdropout)} - 0.14 x_{education(master)}$$

$$+ 0.36 x_{education(collegedropout)} - 0.10 x_{africanamerican(yes)} - 0.08 x_{hispanic(yes)} + 0.51 x_{gender(male)} - 0.01 x_{age} \quad (3)$$

In formula (3), $p(t)$ is the probability of an observation being treated(no smoking bans). $\frac{p(t)}{1-p(t)}$ is called odds.

We take the natural log of odds as the propensity score. −0.81 is the intercept term. 0.67, 1.0, −0.14, 0.36, −0.10, −0.08, 0.51, −0.01 are the coefficients of categorical variables(education, African American, Hispanic and gender). It should be interpreted as the change in log odds for being in different categories. For example, 0.67 represents the change in log odds when the education level is high school rather than college instead while fixing the other variables. We noticed that the education level of college does not exist in formula (3). Because we define all the x variables related to education to be 0 to represent the education level of college. The interpretation for other categorical variables is the same as "education".

**Section 4: The formula to convert log odds into probability**

$$p = \frac{e^y}{1+e^y} \tag{6}$$

In formula (6), p is the probability and y is the log odds. We can use this formula to convert log odds into probability.